\PassOptionsToPackage{table}{xcolor}
\documentclass[10pt,twocolumn,letterpaper]{article}

\usepackage[pagenumbers]{iccv}

\definecolor{blue}{rgb}{0.21,0.49,0.74}
\usepackage[pagebackref,breaklinks,colorlinks,allcolors=blue]{hyperref}
\hypersetup{colorlinks=true, linkcolor=blue, urlcolor=violet}

\usepackage{bbm}
\usepackage{tikz}
\usepackage{listings}
\usepackage{comment}
\usepackage{graphicx}
\usepackage{amsmath}
\usepackage{amssymb}
\usepackage{booktabs}
\usepackage{algorithm}
\usepackage{algpseudocode}
\usepackage[accsupp]{axessibility} 
\usepackage{xfrac}
\usepackage{color}
\usepackage{microtype}
\usepackage{tabularx}
\usepackage{booktabs}
\usepackage{multirow}
\usepackage{blindtext}
\usepackage{xspace}

\usepackage{wrapfig}

\definecolor{dimgray}{rgb}{0.41, 0.41, 0.41}

\usepackage[capitalize]{cleveref}
\crefname{section}{Sec.}{Secs.}
\Crefname{section}{Section}{Sections}
\Crefname{table}{Table}{Tables}
\crefname{table}{Tab.}{Tabs.}

\title{Motion Synthesis with Sparse and Flexible Keyjoint Control}

\author{Inwoo Hwang$^1$\quad Jinseok Bae$^1$\quad Donggeun Lim$^1$\quad Young Min Kim$^{1\dagger}$ \\ \footnotesize{} \\ \normalsize{$^1$ECE, Seoul National University}}

\begin{document}

\twocolumn[{%
\renewcommand\twocolumn[1][]{#1}%
\maketitle

\begin{center}
    \centering
    \captionsetup{type=figure}
    \vspace{-1.0em}
    \includegraphics[width=1.0\linewidth]{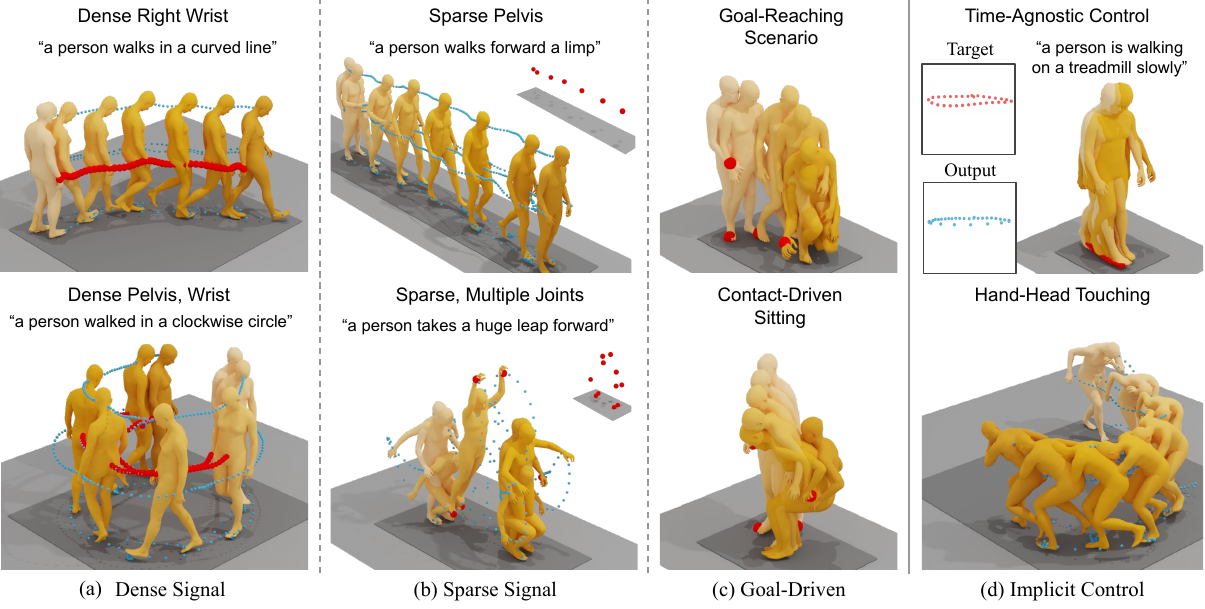}
    \caption{We enable a wide range of practical and controllable motion generation with high quality and precision. Our approach synthesizes natural human motion from \textit{explicit} signals, including (a) \textit{dense} signals with \textit{multiple} joints, (b) \textit{sparse} signals, and (c) \textit{goal}-driven scenarios. Additionally, we generate motion from (d) \textit{implicit} control signals defined via objective functions, including \textit{time-agnostic} control.} 

\label{fig:teaser}
\end{center}
\vspace{1em}
}]

\def\thefootnote{$\dagger$}\footnotetext{Corresponding author}
\def\thefootnote{\arabic{footnote}}

\begin{abstract}

Creating expressive character animations is labor-intensive, requiring intricate manual adjustment of animators across space and time. 
Previous works on controllable motion generation often rely on a predefined set of dense spatio-temporal specifications (e.g., dense pelvis trajectories with exact per-frame timing), limiting practicality for animators.
To process high-level intent and intuitive control in diverse scenarios, we propose a practical controllable motions synthesis framework that respects sparse and flexible keyjoint signals.
Our approach employs a decomposed diffusion-based motion synthesis framework that first synthesizes keyjoint movements from sparse input control signals and then synthesizes full-body motion based on the completed keyjoint trajectories. 
The low-dimensional keyjoint movements can easily adapt to various control signal types, such as end-effector position for diverse goal-driven motion synthesis, or incorporate functional constraints on a subset of keyjoints.
Additionally, we introduce a time-agnostic control formulation, eliminating the need for frame-specific timing annotations and enhancing control flexibility. 
Then, the shared second stage can synthesize a natural whole-body motion that precisely satisfies the task requirement from dense keyjoint movements.
We demonstrate the effectiveness of sparse and flexible keyjoint control through comprehensive experiments on diverse datasets and scenarios. 
Project page: {\footnotesize \url{http://inwoohwang.me/SFControl}}

\end{abstract}
\section{Introduction}
\label{sec:intro}

Motion generation with precise control is critical for the interactivity and realism of character animations in the gaming, virtual reality, and robotics industries.
Traditional animation workflows are labor intensive, requiring animators to specify detailed spatial and temporal constraints manually.
While recent advances synthesize motions that abide by high-level intent signals (e.g., text or audio), scene-aware or interactive tasks may require fine-grained and precise positional control.
In response, emerging methods generate motion guided by spatial control signals.
However, their required inputs are dense frame-wise locations with exact timing, which are challenging to specify manually, severely limiting the practicality of real-world applications.

Instead, we introduce a high-quality motion synthesis framework that abides by sparse and flexible intuitive user control.
Our framework is inspired by the observation that human motion can be intuitively described by the low-dimensional keyjoints movements, such as the end effectors (hands or feet) or body location (root or head).
Our decomposed framework consists of two stages.
The first stage synthesizes keyjoint movements, which precisely adhere to the highly sparse control inputs.
Since the first stage manages low-dimensional data compared to full-body motion, it is highly effectively adapted to respect a wide range of sparse signals.
The subsequent second stage then completes the natural full-body motion from the obtained dense keyjoint trajectories from the first stage.

While it is challenging to generate high-dimensional full-body motion that achieves both naturalness of motion and fidelity to the explicit control signal, the decomposition effectively disentangles controllability from motion quality.
Both stages employ diffusion models and use an imputation strategy while training to allow explicit control~\cite{cohan2024flexible, harvey2022flexiblediffusionmodelinglong, daras2023ambientdiffusionlearningclean}.
In the first stage, the imputed information is randomly selected keyjoint control signals, which earns the ability to adapt to different sets of joints for sparse control signals. 
The imputation for the second stage is the complete keyjoint trajectories, and the second stage can learn to generate a high-quality full-body motion with complete keyjoint signal.
Both stages can condition on the text descriptions or action labels to maintain the natural semantic intent of the overall motion sequence.

We can flexibly define a desired objective function on a subset of keyjoints, achieving user-friendly control.
For example, the animator can only specify joint trajectories without time information, which could not be processed with previous approaches.
We can allow time-agnostic control on the keyjoint trajectory by formulating a differentiable optimization objective on the spatial trajectory regardless of the time stamp. 
Additionally, we can efficiently tackle goal-driven motion generation, where the goal pose serves as a highly sparse spatial constraint.
We demonstrate that our generated motion can perform diverse tasks with different sets of goal specifications in a unified setup without dataset-specific training: reaching target hand positions~\cite{araujo2023circle}, climbing with rock constraints~\cite{yan2023cimi4d}, and sitting with hand control~\cite{zhang2022couch}. 

Our contributions can be summarized as follows:
(1)	We propose a decomposed, diffusion-based framework for controllable motion synthesis that accurately follows sparse control signals while maintaining high-quality motion.
(2)	Our method effectively respects highly sparse control signals by concentrating on low-dimensional keyjoints movements at first stage within decomposed framework
(3)	We formulate diverse keyjoint constraints that enhance user controllability, such as time-agnostic control, goal-driven motion synthesis, and various constraint functions.

\section{Related Works}
\label{sec:related}

\paragraph{Human Motion Generation with Semantic Input}

Recent advances in data-driven human motion synthesis have enabled the generation of highly realistic and diverse human motions. Modern generative modeling techniques—such as generative adversarial networks (GANs)~\cite{xu2022actformerganbasedtransformergeneral}, variational autoencoders (VAEs)~\cite{Guo_2022_CVPR}, token-based models~\cite{guo2023momask}, and diffusion models~\cite{tevet2023human}—have successfully captured the complexity of human motion dynamics. Many works focused on synthesizing motions that adapt to high-level semantic conditions (e.g., action categories~\cite{guo2020action2motion, petrovich2021action, tevet2023human, chen2023executing, hoang2024motionmix}, text descriptions~\cite{chuan2022tm2t, guo2023momask, zhang2023generating, yuan2023physdiff, bae2025moreimprovingmotiondiffusion, Guo_2022_CVPR, tevet2023human, zhang2022motiondiffuse, petrovich2022temos, lin2023comes, petrovich24stmc, snapmogen2025}, audio~\cite{lee2019dancing, gong2023tm2d, alexanderson2023listen, siyao2022bailando, tseng2022edge, li2021learn, Qi2023diffdance, }) and surrounding environments (e.g., interacting objects~\cite{xu2023interdiff, ghosh2022imos, li2023object, Zhao2022coins, bhatnagar22behave, kulkarni2023nifty, taheri2020grab, bae2023pmp, li2023controllable} or scenes~\cite{TRUMANS, humanise, LINGO, mammos, lee2023lama, DIMOS, tesmo,cen2024text_scene_motion, SceneDiffuser, AffordMotion, sui2025surveyhumaninteractionmotion, Hwang_2025_CVPR}). These methods bridge the gap between user intent and complex motion generation, enabling intuitive applications for animation.

\paragraph{Motion Synthesis with Full-Body Pose Input}
In addition to high-quality motion generation from high-level intents (e.g., text prompts), allowing spatially precise control is critical for practical applications, particularly in interactive animation workflows. A foundational task in this domain is motion in-betweening~\cite{arikan2002interactive, beaudoin2008motion, kovar2023motion, harvey2018recurrent, zhang2018data, harvey2020robust, harvey2020robust, duan2021single, qin2022motion, oreshkin2023motion,kim2022conditional, hwang2025scenemimotioninbetweeningmodeling}, which synthesizes plausible transitions between keyframes defined as full-body information. 
Early works explored interpolation using splines such as Bézier curves~\cite{708559, 10.1145/1073204.1073313, 6909567}, but these methods often lacked the flexibility to model complex human motion. Recent neural approaches have improved robustness: Qin et al.~\cite{qin2022motion} proposed a two-stage deterministic network for motion completion, while CondMDI~\cite{cohan2024flexible} leveraged generative capabilities of diffusion models and extended their abilities to incorporate text instructions.

\paragraph{Motion Synthesis with Joint-Level Control}
Beyond keyframe interpolation, recent methods focus on joint-level spatial control for advanced motion synthesis. 
PriorMDM~\cite{shafir2023human} pioneered joint-aware generation by fine-tuning a pre-trained motion diffusion model MDM~\cite{tevet2023human}. GMD~\cite{shafir2023human} introduces classifier-free guidance to diffusion models to enable root joint trajectory control, and OmniControl~\cite{xie2023omnicontrol} generalized the control capability to arbitrary joints via ControlNet~\cite{zhang2023adding} modules with spatial guidance at inference. MotionLCM~\cite{motionlcm} further incorporated ControlNet into a motion latent space with a consistency model.
Other approaches explored latent space optimization: TLControl~\cite{wan2023tlcontrol} trained part-based discrete latent codes and performed inference-time code optimization, while ControlMM~\cite{pinyoanuntapong2024controlmm} exploited spatial guidance into a masked motion model via token optimization. DNO~\cite{karunratanakul2023dno} optimized diffusion noise via a differentiable objective function defined through global joint position. 
A recent concurrent work, Studer \etal~\cite{FactorizedDiffusion} introduced a factorized diffusion approach with Bézier curve parameterization, though it omits text conditioning.

\begin{figure*}[t!]
\centering
\includegraphics[width=1.0\linewidth]{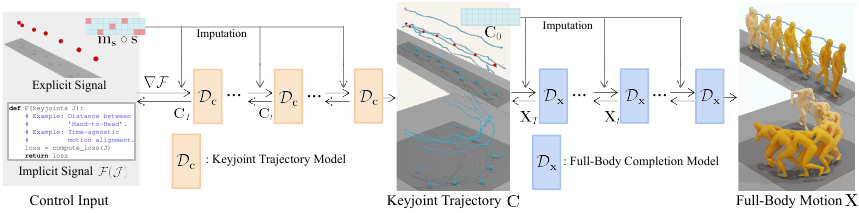}
\caption{From control input, we first synthesize the keyjoint trajectory $\mathbf{C} \in R^{N\times d}$ as an intermediate representation. For \textit{explicit} control, the signal $\mathbf{s}$ is combined with the binary indicator $\mathbf{m_s} \in \{0, 1\}^{N\times d}$. 
We impute $\mathbf{m_s} \circ \mathbf{s}$ to the keyjoint trajectory and denoise it through keyjoint trajectory model $\mathcal{D}_{\mathbf{c}}$ at each diffusion step. 
For \textit{implicit} control, an objective function $\mathcal{F(\cdot)}$ is defined over the keyjoint $\mathcal{J}$ and optimized thorough diffusion latent optimization. The resulting keyjoint trajectory $\mathbf{C}$ is then imputed into the full-body motion completion model $\mathcal{D}_{\mathbf{x}}$ to synthesize the final full-body motion $\mathbf{X} \in R^{N\times D}$.}
\label{fig:architecture}
\vspace{-1.0em}
\end{figure*}

Despite these advancements, most existing methods rely on dense control signals (e.g., full-sequence pelvis trajectories), and the performance degrades under sparse or partial joint constraints. They also assume strict temporal alignment between control signals and generated motions, limiting applicability to real-world scenarios with asynchronous or ambiguous timing. Addressing these gaps remains essential for practical deployment in interactive animation systems.
\section{Motion Synthesis with Keyjoint Control}
\label{sec:methods}

Given joint-level control signals, our goal is to generate natural and precise full-body character motion while satisfying the input constraints. 
We define keyjoints set $\mathcal{J}$, as the end-effectors (hands and feet) and critical body locations (pelvis (root) and head), which collectively capture the essential dynamics of full-body human motion.
The control signals are either the \textit{explicit} positions with corresponding timing inputs or an \textit{implicit} function of a set of keyjoints.

Inspired by the observation that human motion can be effectively described through keyjoint dynamics, we propose a decomposed diffusion framework for controllable motion synthesis. 
The pipeline operates in two stages: we first synthesize keyjoint trajectories that precisely satisfy the input control signal (Sec.~\ref{sec:Key-Diffusion}), and subsequently synthesize full-body motion conditioned on these trajectories (Sec.~\ref{sec:Full-Body-Diffusion}).
The keyjoint movements $\mathbf{C}$ serve as a low-dimensional intermediate representation that contains the essential dynamics of human motion $\mathbf{X}$, thereby decomposing the motion synthesis problem into simplified stages.
The overall process is illustrated in Figure~\ref{fig:architecture}.

\paragraph{Motion Representation}
The keyjoint trajectory representation is denoted as $\mathbf{C} =\{ \mathbf{c}^n\}_{n=1}^{N}$, 
where each $\mathbf{c}^n \in \mathbb{R}^d$ encodes the kinematic state of the $n$-th frame's keyjoint. Here, $d$ denotes the dimensionality of the joint’s feature representation, including the global positions of keyjoints $J \in \mathbb{R}^{6 \times 3}$ and a root's y-axis rotation $\gamma \in \mathbb{R}^1$.
Full-body motion is represented as $\mathbf{X} = \{ \mathbf{x}^n\}_{n=1}^{N}$, 
where each $\mathbf{x}^n \in \mathbb{R}^D$ represents a full-body pose with dimensionality $D \gg d$. We adopt the HumanML3D~\cite{Guo_2022_CVPR} representation and convert the root information into global coordinate system. This conversion aligns with the methodology in~\cite{karunratanakul2023gmd}, where the feature dimension is $D=263$.

During the first stage, we synthesize the movement of the keyjoint set $\mathcal{J}$ in the global coordinate. 
The resulting keyjoint trajectory $\mathbf{C} =\{ \mathbf{c}^n\}_{n=1}^{N}$ precisely reflects joint-level constraints, while providing essential and sufficient information for predicting full-body movement.
We then convert $\mathbf{C}$ to be in the coordinates relative to the root to align with the full-body representation.
In the next stage, we synthesize the complete full-body motion $\mathbf{X} = \{ \mathbf{x}^n\}_{n=1}^{N}$.

\subsection{Keyjoint Trajectory Model}

\label{sec:Key-Diffusion}

We primarily train the diffusion model using a conditional data distribution $p(\mathbf{C}_0, l)$, where $\mathbf{C}_0$ is a keyjont trajectory and $l$ denotes an additional condition (e.g., text or action label).
During training a diffusion model,  noise is progressively added to the data at timestep $t$ within the total diffusion steps $T$.
The noisy trajectory sample at timestep $t$ is generated by $\mathbf{C}_t = \sqrt{\bar{\alpha}_t} \mathbf{C}_0 +  \sqrt{1 - \bar{\alpha}_t}\boldsymbol{\epsilon}$, where $\bar{\alpha}_t$ is determined by diffusion noise scheduling and $\epsilon$ is noise sampled from i.i.d. Gaussian distribution.

The goal of this stage is to synthesize keyjoint movements $\mathbf{C}$ that satisfy the given keyjoint control signal $\mathbf{s}$.
The keyjoint control signal $\mathbf{s}$ is either explicit \textit{xyz} positions of selected keyjoints, indicated by a binary indicator $\mathbf{m_s} \in \{0, 1\}^{N\times d}$, or an implicit objective function $\mathcal{F(\cdot)}$ defined over the keyjoint set $\mathcal{J}$.
Below, we further elaborate the formulations incorporating explicit control signal (Sec.~\ref{sec:Explicit-Control}) and implicit control signal (Sec.~\ref{sec:Time-Agnostic}).

\subsubsection{Motion Generation with Explicit Keyjoint Control}
\label{sec:Explicit-Control}

The input for explicit control signal is provided by 3D \textit{xyz} locations with exact time stamps.
Therefore, we can employ imputation scheme to process sparse selections in both time and joint indices (Figure~\ref{fig:architecture}, top left).

During training, the denoising network incorporates an imputation strategy with a randomly generated mask, earning the flexibility to handle arbitrary control signals.
We select a timestep $t$ from the uniform distribution over $[1, T]$ and generate a random spatio-temporal mask $\mathbf{m_C} \in \{0, 1\}^{N\times d}$.
This mask $\mathbf{m_C}$ indicates which joints will be used for conditional motion synthesis.
We then impute the noisy sample $\mathbf{C}_t$ with the corresponding values from the clean sample $\mathbf{C}_0$:
\[
\mathbf{C}_t' = \mathbf{m_C} \circ \mathbf{C}_0 + (1 - \mathbf{m_C}) \circ \mathbf{C}_t,
\]
where $\circ$ denotes element-wise multiplication.
Next, we concatenate the imputed sample $\mathbf{C}_t'$ with the indicator mask $\mathbf{m_C}$ to form the combined feature:
\[
\tilde{\mathbf{C}}_t = \langle \mathbf{C}_t', \mathbf{m_C} \rangle.
\]
Finally, we train a keyjoint trajectory denoiser $\mathcal{D}_{\mathbf{c},\theta}$ by minimizing the following loss function:
\[
\label{eq:simple_loss_key}
\mathcal{L}_{\text{simple},\mathbf{c}} = \mathbb{E}_{\mathbf{C}_0 \sim p(\mathbf{C}_0|l), t \sim [1, T]} \left[ \left\| \mathbf{C}_0 - \mathcal{D}_{\mathbf{c},\theta}(\tilde{\mathbf{C}}_t, t, l) \right\|_2^2 \right].
\]

At inference, we synthesize the keyjoint trajectory $\mathbf{C}_0$ from the explicit control signal $\mathbf{s}$, along with the binary indicator $\mathbf{m_s} \in \{0, 1\}^{N\times d}$.
We start inference by imputing noisy samples with  $\mathbf{s}$, similar to the training process
\[
\mathbf{C}_t' = \mathbf{m_s} \circ \mathbf{s} + (1 - \mathbf{m_s}) \circ \mathbf{C}_t.
\]
Then, we use the concatenated temporal features $
\tilde{\mathbf{C}}_t = \langle \mathbf{C}_t', \mathbf{m_s} \rangle
$ and generate motion sequences using a conditional diffusion process with classifier-free guidance.
\[
\hat{\mathbf{C}}_0 = \mathbf{w} \cdot \mathcal{D}_{\mathbf{c},\theta}(\tilde{\mathbf{C}}_t, t, l) + (1 - \mathbf{w}) \cdot \mathcal{D}_{\mathbf{c},\theta}(\tilde{\mathbf{C}}_t, t, \emptyset),
\]
where $\mathbf{w}$ is the guidance weight at inference~\cite{ho2022classifierfreediffusionguidance}.

\paragraph{Goal-Driven Motion Synthesis}
\label{sec:Goal-Driven}
One of highly sparse user input for explicit control is specifying the goal position of a selected joint, such as a hand.
Goal-driven motion synthesis aims to complete the transition motion starting from a given pose and reaching a target goal position defined for a specific keyjoint.
To tailor our approach for the task, we adapt the binary indicator for the control signal $\mathbf{m_C}$ during training to include only the start pose and the target control joint at the final frame.
We train a unified network on three different task configurations. 
In addition to the action label for the goal task, the network is conditioned on the continuous body shape encoding to account for the spatial relationship between the body size and the specific goal locations.
The detailed training procedures and implementation details are provided in the supplementary material.

\subsubsection{Motion Generation with Implicit Keyjoint Control}
\label{sec:Time-Agnostic} 

In addition to explicit spatial control, we can synthesize motion that satisfies high-level constraints, such as avoiding obstacles.
Specifically, our framework can leverage implicit control signals, which are defined as differentiable objective functions $\mathcal{F(\cdot)}$ defined over the keyjoint set $\mathcal{J}$.
The functional objectives can impose flexible and implicit constraints.
For example, we can generate motions that demonstrate 'Hand-to-Head,' by defining the objective function to be the distance between the hand and head, and minimizing it.

We maintain the same model trained in the previous stage (Sec.~\ref{sec:Explicit-Control}), with modifications applied only at inference time.
Since the implicit control signal does not contain frame-wise index, we cannot employ the imputation as described in \cref{sec:Explicit-Control}.
Instead, we refine motion sequences by back-propagating the derivatives of the objective function while iteratively optimizing the diffusion noise in the latent space~\cite{karunratanakul2023dno}.
The motion sequence satisfies the desired functional criteria while maintaining natural movement.

\paragraph{Time-Agnostic Motion Control}
Our functional constraint can allow trajectory input without exact per-frame timestamps, namely \textit{time-agnostic motion control}.
One critical limitation of explicit spatial motion control is the requirement for the exact frame-wise timing.
Such information is highly challenging for an animator to accurately annotate, limiting their practicality in real-world scenarios.

We can formulate a flexible input constraint based on arc-length reparametrization.
We assume the input composed of only a start time  $t_{0}$, an end time $t_{1}$, and the desired 3D target trajectory $\mathcal{T}$ of the specific keyjoint  over time interval $[t_{0}, t_{1}]$.
Given a target constraint $(\mathcal{T}, t_{0}, t_{1})$, our method enforces geometric consistency against the generated keyjoint trajectories $\mathbf{C} =\{ \mathbf{c}^n\}_{n=1}^{N} \in \mathbb{R}^{N \times 3}$ through time-agnostic trajectory alignment process.
We first extract the segment $\mathbf{C}_{\text{seg}}$ of the specified keyjoint during the time window \([t_0, t_1]\) from the generated trajectory  $\mathbf{C} =\{ \mathbf{c}^n\}_{n=1}^{N}$:
\[
\mathbf{C}_{\text{seg}} = \{ \mathbf{c}^n \}_{n=t_0}^{t_1} \in \mathbb{R}^{L \times 3}, \quad L = t_1 - t_0 + 1.
\]
Then, we calculate cumulative  arc lengths $\{s_i\}_{i=1}^L$ of $\mathbf{C}_{\text{seg}}$:
\[
s_i = \begin{cases} 
    0, & i=1 \\
    s_{i-1} + \|\mathbf{c}^{t_0+i-1} - \mathbf{c}^{t_0+i-2}\| & i=2,\dots,L 
\end{cases}
\]
where $s_L$ is the total arc length. 
After that, we obtain a cubic spline parametrized by the temporal intervals $\mathcal{S}(s): [0, s_L] \mapsto \mathbb{R}^3$, which approximates $\mathbf{C}_{\text{seg}}$, with a differentiable process~\cite{kidger2022torchcubicspline}.
To obtain a temporally invariant representation, we uniformly sample  $L$ points on the spline and construct a version with arc-length parameterization 
$\mathbf{C}_{\text{seg}}^{\text{res}} = \{\mathcal{S}(\tilde{s}_k)\}_{k=1}^{L} \in \mathbb{R}^{L \times 3}$:
\[
\tilde{s}_{k} = \frac{k-1}{L-1} \cdot s_L, \quad k \in \{1, \dots, L\}.
\]
Similarly, we uniformly sample the target trajectory $\mathcal{T}$ to obtain $\mathcal{T}^{\text{res}} \in \mathbb{R}^{L \times 3}$.
The \textit{time-agnostic motion alignment loss} combines geometric and scale consistency:
\begin{equation}
\label{eq:time_agnostic}
\mathcal{L}_{\text{align}} = 
\underbrace{\left\| \mathbf{C}_{\text{seg}}^{\text{res}} - \mathcal{T}^{\text{res}} \right\|}_{\text{Traj Align.}} + 
\lambda_l \cdot 
\underbrace{\left| s_L - L(\mathcal{T}) \right|}_{\text{Length Const.}},
\end{equation}
where $L(\mathcal{T})$ is the total length of $\mathcal{T}$. The first term ensures geometric consistency by aligning the uniform samples of the generated trajectory with the target trajectory, independent of temporal information, while the second ensures consistent motion scale. 
By optimizing the diffusion latent noise~\cite{karunratanakul2023dno} with the proposed objective in \cref{eq:time_agnostic}, our method enables control over the keyjoint trajectory without requiring exact per-frame time stamps.
\subsection{Full-Body Motion Completion Model}
\label{sec:Full-Body-Diffusion}

The second stage generates a natural full-body motion sequence $\mathbf{X} = \{ \mathbf{x}^n \} \in \mathbb{R}^{N \times D}$ that follows the dense keyjoint trajectory $\mathbf{C}$ from the first stage. 
We train the conditional diffusion model from the joint data distribution $p(\mathbf{C}_0, \mathbf{X}_0, l)$.

\paragraph{Training}

During training, we add noise corresponding to diffusion timestep $t$ and obtain noisy full-body motion $\mathbf{X}_t = \sqrt{\bar{\alpha}_t} \mathbf{X}_0 +  \sqrt{1 - \bar{\alpha}_t}\boldsymbol{\epsilon}$.
Similar to the keyjoint model training, we impute the clean keyjoint information $\mathbf{C}_0$ into $\mathbf{X}_t$. 
Specifically, given a noisy full-body motion sample $\mathbf{X}_t$ at diffusion timestep $t$, we replace its corresponding keyjoint component with the ground truth from $\mathbf{C}_0$:
\[
\mathbf{X}_t^\text{keyjoints} = \mathbf{C}_0.
\]
This imputation preserves the provided keyjoint movement throughout the diffusion process, effectively guiding the full-body motion synthesis.

The full-body denoiser $\mathcal{D}_{\mathbf{x},\theta}$ is then trained to reconstruct the original full-body motion $\mathbf{X}_0$ from the noisy input $\mathbf{X}_t$, conditioned on $l$. The training objective is given by:
\[
\label{eq:simple_loss_full}
\mathcal{L}_{\text{simple},\mathbf{x}} = \mathbb{E}_{\mathbf{X}_0 \sim p(\mathbf{X}_0|\mathbf{C}_0, l), t \sim [1, T]} \left[ \left\| \mathbf{X}_0 - \mathcal{D}_{\mathbf{x},\theta}(\mathbf{X}_t, t, l) \right\|_2^2 \right].
\]

\paragraph{Inference}

At inference, we synthesize the full-body motion $\mathbf{X}_0$ given estimated keyjoint trajectories $\hat{\mathbf{C}}_0$.
We start inference by imputing noisy samples with  $\hat{\mathbf{C}}_0$, similar to the training process
\[
\mathbf{X}_t^\text{keyjoints} = \hat{\mathbf{C}}_0
\]
and sample via a conditional diffusion process:
\[
\hat{\mathbf{X}}_0 = \mathbf{w} \cdot \mathcal{D}_{\mathbf{x},\theta}(\mathbf{X}_t, t, l) + (1 - \mathbf{w}) \cdot \mathcal{D}_{\mathbf{x},\theta}(\mathbf{X}_t, t, \emptyset),
\]
where $\mathbf{w}$ is the guidance weight.

\section{Experiments}
\label{sec:results}

We first evaluate our controllable motion synthesis framework in explicit control signal where exact time and locations for keyjoints are given (Sec.~\ref{sec:exp_sparse}).
Additionally, we present results from implicit motion control via differentiable objectives defined with keyjoints (Sec.~\ref{sec:exp_function}).
After analyzing full-body motion completion model (Sec.~\ref{sec:exp_full}), we further investigate diverse aspects of our keyjoint control pipeline (Sec.~\ref{sec:perfom_analysis}).
We provide detailed descriptions on the datasets and baselines in the supplementary material.

\paragraph{Implementation Details}

Our pipeline is implemented in PyTorch~\cite{pytorch} and trained on a single NVIDIA RTX 2080 GPU.
We adopt the U-Net architecture from~\cite{karunratanakul2023gmd}.
Further details on model architecture and diffusion hyperparameters are provided in the supplementary material.

\subsection{Explicit Keyjoint Control}
\label{sec:exp_sparse}

We begin our evaluation using the HumanML3D dataset, which contains text annotations paired with motion data captured at 20 FPS. We compare our method with state-of-the-art approaches for joint-level motion synthesis, selecting those with publicly available code.

For our experiments, we evaluate our framework under a highly sparse control setting.
We set the control signal interval to 
$r=30$ frames (i.e., one frame every 1.5 seconds) and further randomly select 50$\%$ out of six  keyjoints at each frame. This corresponds to only 0.454$\%$ of the total joint signals contributing to the full-body motion, calculated as $
(1/30) \times (6/22) \times (1/2) = 0.00454$,
where $1/30$ accounts for the frame sampling rate, $6/22$ represents the ratio of key joints to total joints, and $1/2$ reflects the random retention rate of control signals.

We evaluate our framework across spatial control accuracy, motion realism, text fidelity, and diversity.
For spatial control accuracy, we assess motion controllability via \textit{Control Error} by computing the Euclidian distance between generated motions and the provided spatial targets. Motion quality is evaluated through two metrics: \textit{Frechet Inception Distance (FID)} which measures distribution alignment between ground truth and generated, and the \textit{Foot Skating Ratio} qunatifies the artifact of foot's movement. \textit{R-Precision} measures text-motion alignment and \textit{Diversity} measures the variability within the generated motion.

\Cref{tab:experiments_sparse_control} presents the quantitative results.
Compared to other baselines, our model achieves lower \textit{FID} scores, lower \textit{control error}, while maintaining \textit{diversity} close to the ground truth, demonstrating superior motion quality and precision under sparse control conditions.
These results highlight the effectiveness of our approach in generating high-quality, diverse, and precise motions from sparse control input.
The two stage diffusion framework is not real-time as it involves multiple  sampling steps (DDPM).
We also test a variation of our pipeline using the DDIM~\cite{song2020denoising} sampling strategy, which reduces sampling time, with sampling steps of five (DDIM5) and ten (DDIM10).
Notably, our method consistently outperforms baselines even when employing the faster DDIM sampling strategy. 
We provide more analysis of runtime performance in \cref{sec:perfom_analysis}.

\begin{table}[ht!]
    \centering
    \renewcommand{\arraystretch}{1.1}
\resizebox{1.0\linewidth}{!}{
\begin{tabular}{lccccc}
\toprule
\multirow{2}{*}{\centering Method} & \multirow{2}{*}{\centering FID~$\downarrow$} & \multicolumn{1}{p{1.41cm}}{\centering Control Err. ($m$)~$\downarrow$} & \multicolumn{1}{p{1.65cm}}{\centering R-precision (Top-3)~$\uparrow$} & \multirow{2}{*}{\centering Div.~$\rightarrow$} & \multicolumn{1}{p{1.37cm}}{\centering Foot Skating~$\downarrow$} \\

\midrule
\rowcolor{gray!15}
Real                                   &  0.002 & 0.000 & 0.797 & 9.503 & 0.000 \\
\midrule
CondMDI~\cite{cohan2024flexible}       &  0.498 & 0.507 & 0.631 & 9.013 & 0.099\\
OmniControl~\cite{xie2023omnicontrol}  &  0.689 & 0.111 & \textbf{0.689} & \underline{9.381} & 0.091 \\
TLControl~\cite{wan2023tlcontrol}      &  4.637 & 0.128 & 0.473 & 8.078 & \textbf{0.058} \\
MotionLCM~\cite{motionlcm}             &  1.013 & 0.418 &\underline{ 0.676} & 8.722 & 0.141 \\
\midrule
Ours (DDIM5)                          &  0.338 & 0.037 & 0.655 & 9.356 & 0.065 \\
Ours (DDIM10)                          &  \underline{0.254} & \underline{0.037} & 0.673 & \textbf{9.621} & 0.063 \\
Ours                                   &  \textbf{0.224} & \textbf{0.036} & 0.673 & 9.674 &\underline{0.061} \\
 \bottomrule
\end{tabular}
}
\caption{Quantitative evaluation of the sparse joint control. We validate our framework under a highly sparse control setting, which uses only $0.454\%$ of total joint signal as control input. \textbf{Bold} represents the best value, and \underline{underlined} represents the second-best.
}
\vspace{-1em}
\label{tab:experiments_sparse_control}
\end{table}

\begin{figure}[ht!]
\centering
\includegraphics[width=0.9\linewidth]{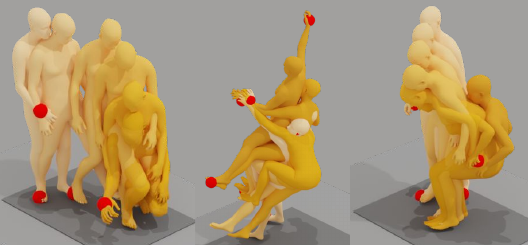}
\caption{Qualitiative results of goal-driven motion scenarios, demonstrating \textit{reaching target hand positions}, \textit{climbing with rock constraints}, and \textit{sitting with hand control}, respectively.}
\label{fig:goal_driven}
\end{figure}

\subsubsection{Goal-Driven Motion Synthesis}

As a practical extension of sparse explicit control, 
we adapt the framework for goal-driven motion synthesis by training a unified network that integrates multiple motion scenarios. In each scenario, control signals are defined by an initial pose and target control joints at the final frame. Specifically, \textit{reaching target hand positions}~\cite{araujo2023circle} focuses on controlling the right-hand position, \textit{climbing with rock constraints}~\cite{yan2023cimi4d} requires coordination of both hands and feet, and \textit{sitting with hand control}~\cite{zhang2022couch} necessitates precise control of both hands at the final frame, as illustrated  in \Cref{fig:goal_driven}. Our unified network is designed to handle multiple motion dynamics and control settings within a single framework.

We benchmark our approach against a single-stage version that does not detach keyjoint movements during training. 
To assess goal-reaching accuracy, we measure the average Euclidean distance between the final generated poses and the target keyjoint positions (\textit{Distance to Goal}). And also measure the deviation from the input starting full-body pose (\textit{Distance to Start}), and \textit{Foot Skating}.
As shown in \cref{tab:goal_driven}, our decomposed pipeline demonstrates substantial advantages by prioritizing on low-dimensional key joint movements in the first stage.
Ours can satisfy the start and the goal positions more precisely with minimal foot skating artifacts.
Furthermore, our model maintains reasonable performance even with limited task-specific data: For the \textit{sitting with hand control} and \textit{climbing with rock constraints} scenarios, we train on only 160 and 156 motion segments, respectively, yet achieve competitive results. 
We further analyze generalization performance to different control inputs in Sec.~\ref{sec:perfom_analysis}.

\begin{table}[h!]
    \centering
    \renewcommand{\arraystretch}{1.0} 
    \resizebox{0.95\linewidth}{!}{
        \begin{tabular}{llcccc}
            \toprule
            \multicolumn{1}{p{0.6cm}}{} &
            \multirow{2}{*}{\centering Scenario} & 
            \multicolumn{1}{p{1.7cm}}{\centering Dist. to Goal ($m$)~$\downarrow$} & 
            \multicolumn{1}{p{1.7cm}}{\centering Dist. to Start ($m$)~$\downarrow$} & 
            \multicolumn{1}{p{1.7cm}}{\centering Foot Skating~$\downarrow$} \\
            \midrule
            \multirow{4}{*}{\rotatebox[origin=c]{90}{Ours w/o}}
            \multirow{4}{*}{\rotatebox[origin=c]{90}{decomp.}}
             & All      & 0.206 & 0.116 & 0.057   \\
             & Reaching & 0.141 & 0.061 & 0.051   \\
             & Climbing & 0.529 & 0.409 & 0.079  \\
             & Sitting  & 0.327 & 0.126 & 0.092   \\
            \midrule
            \multirow{4}{*}{\rotatebox[origin=c]{90}{}}
            \multirow{4}{*}{\rotatebox[origin=c]{90}{Ours}}
             & All      & \textbf{0.093} & \textbf{0.065} & \textbf{0.047}   \\
             & Reaching & \textbf{0.054} & \textbf{0.043} & \textbf{0.048}   \\
             & Climbing & \textbf{0.288} & \textbf{0.185} & \textbf{0.046}   \\
             & Sitting  & \textbf{0.156} & \textbf{0.064} & \textbf{0.050}  \\
            \bottomrule
        \end{tabular}
    }
    \caption{Quantitative evaluation on the \textit{goal-driven} scenarios. We train a unified network across three different tasks and evaluate it separately for each task as well as collectively.}
    \vspace{-0.4em}
    \label{tab:goal_driven}
\end{table}

\subsection{Implicit Keyjoint Control}
\label{sec:exp_function}

Additionally, we evaluate our approach under implicit control settings, where constraints are defined through differentiable functions, including a time-agnostic control scenario. To assess its effectiveness, we adapt evaluation tasks from \cite{liu2024programmable}, such as generating hand-to-head contact movements and walking in narrow spaces.
To quantify performance, we report \textit{Constraint Error}, which measures the degree to which user-defined constraints are satisfied, and the \textit{Critic Score}~\cite{motioncritic2025}, which assesses motion naturalness and quality based on human perceptual judgments. For each scenario, we randomly sample 10 instances for evaluation.

We compare our approach against both the original implementation of \cite{karunratanakul2023dno}, which directly generates full-body motion, and our single-stage variant  (referred to as ``Ours w/o decomposed"). 
As presented in \Cref{tab:joint_loss}, our decomposed pipeline consistently improves constraint error while enhancing motion quality.
By applying objectives at the keyjoint trajectory level rather than at the full-body motion synthesis stage, our approach better satisfies objective constraints while preserving the realism of the full-body motion after completion.

\begin{table}[t!]
    \centering
    \renewcommand{\arraystretch}{1.08}
    \resizebox{1.0\linewidth}{!}{
        \begin{tabular}{llccc}
            \toprule
            \multirow{2}{*}{Task} &
            \multirow{2}{*}{\centering Method} & 
            \multicolumn{1}{p{1.3cm}}{\centering Constraint Error~$\downarrow$} & 
            \multicolumn{1}{p{1.6cm}}{\centering Critic Score~$\uparrow$} & 
            \multicolumn{1}{p{1.6cm}}{\centering Foot Skating~$\downarrow$} \\
            \midrule
            \multirow{3}{*}{\rotatebox[origin=c]{90}{Walking}}
            \multirow{3}{*}{\rotatebox[origin=c]{90}{Narrow}}
            \multirow{3}{*}{\rotatebox[origin=c]{90}{}}
             & DNO~\cite{karunratanakul2023dno} & 0.0101 & -2.024 & 0.142  \\
             & Ours w/o decomp. & 0.0093 & -2.192 & 0.090  \\
             & Ours & \textbf{0.0084} & \textbf{0.253}  & \textbf{0.066}  \\
            \midrule
            \multirow{3}{*}{\rotatebox[origin=c]{90}{Hand-to}}
            \multirow{3}{*}{\rotatebox[origin=c]{90}{-Head}}
            \multirow{3}{*}{\rotatebox[origin=c]{90}{}}
             & DNO~\cite{karunratanakul2023dno} & 0.0084 & -6.303 & 0.064   \\
             & Ours w/o decomp. & 0.0093 & -5.842 & 0.059  \\
             & Ours & \textbf{0.0036} & \textbf{-2.505}  & \textbf{0.045}   \\
             \midrule
             
             \multirow{4}{*}{\rotatebox[origin=c]{90}{Time-}}
            \multirow{4}{*}{\rotatebox[origin=c]{90}{Agnostic}}
            \multirow{4}{*}{\rotatebox[origin=c]{90}{}}
             & DNO~\cite{karunratanakul2023dno}   & 0.0412 & 0.246 & 0.061  \\
             & Ours w/o decomp.                   & 0.0581 & 0.234 & 0.058  \\
             & Ours w/o time-agnostic             & 0.4827 & 0.213 & 0.097  \\
             & Ours                               & \textbf{0.0256} & \textbf{0.398} & \textbf{0.048}  \\
            \bottomrule
        \end{tabular}
    }
    \caption{Quantitative evaluation on different objective defined task scenarios.} 
    \vspace{-1.4em}
    \label{tab:joint_loss}
\end{table}

\begin{figure}[h!]
\centering
\includegraphics[width=0.99\linewidth]{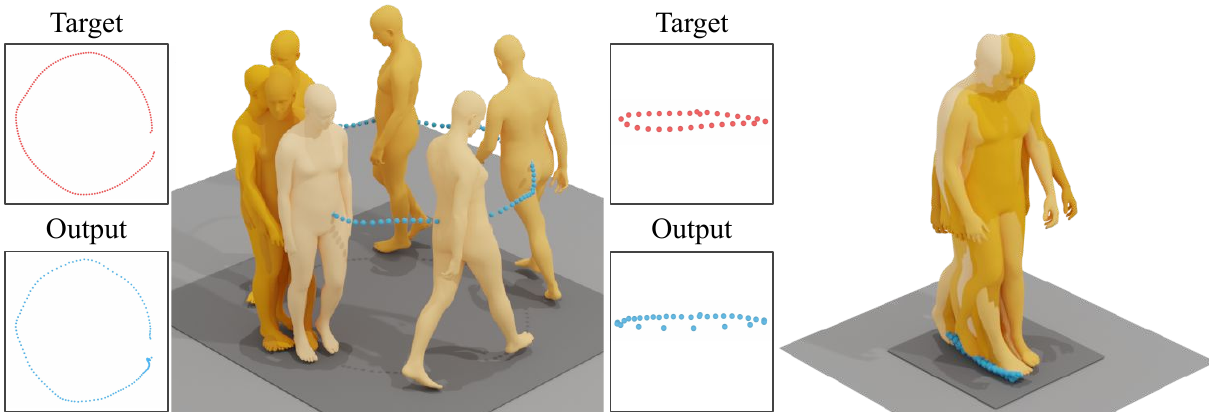}
\caption{Example of time-agnostic trajectory target input and synthesized motion from time-agnostic control.}
\vspace{-0.6em}
\label{fig:time_agnostic}
\end{figure}

To further validate the effectiveness of our \textit{time-agnostic motion control} framework, we compare our proposed implicit control approach with an explicit  approach as ``Ours w/o time-agnostic''.
Specifically, for a trajectory $\mathcal{T}$ where only the start and end times, $t_0$ and $t_1$, are provided, we sample points along the trajectory to have the same length between them. 
Then we assign timestamps with a uniform time interval between $t_0$ and $t_1$ on the sampled points and provide the locations as explicit control signal.
Table~\ref{tab:joint_loss} shows that the explicit model without exact time information struggles to accurately follow the given trajectory. 
In contrast, our implicit formulation for time-agnostic control effectively overcome the limitation with our dedicated time-agnostic motion alignment objective.
Figure~\ref{fig:time_agnostic} visualizes the input trajectory $\mathcal{T}$, which lacks timing annotations, alongside the corresponding trajectory generated using our time-agnostic control. Notably, our method produces target trajectory with non-uniform intervals, effectively capturing velocity variations.
Overall, our decomposed pipeline offers flexibility under implicit control, including time-agnostic motion alignment mechanism.

\subsection{Full-Body Motion Completion Model}
\label{sec:exp_full}

We also evaluate the performance of our full-body motion completion model to assess how accurately the synthesized full-body motion follows the key joint movements.
In ~\cref{tab:stage2_verification}, we measure the \textit{Mean Euclidean Distance Error} between input keyjoint trajectories and the final full-body motion derived from different keyjoint sources: ground-truth and synthesized keyjoints from keyjoint trajectory model. The results demonstrate that our full-body motion precisely adheres to the given key joint movements and that the synthesized keyjoint trajectories remain faithful to those of the ground truth.

\begin{table}[h!]
    \centering
    \renewcommand{\arraystretch}{1.0}
    \resizebox{0.95\linewidth}{!}{
        \begin{tabular}{lcc}
            \toprule
            \multicolumn{1}{p{2.6cm}}{KeyJoint Source} & 
            \multicolumn{1}{p{2.6cm}}{\centering Ground Truth} & 
            \multicolumn{1}{p{2.6cm}}{\centering Synthesized} \\
            \midrule
              Mean Distance Error (m) &  0.0082 & 0.0115 \\
            \bottomrule
        \end{tabular}
    }
    \caption{Quantitative evaluation of full-body motion completion model on different keyjoint source.}
    \label{tab:stage2_verification}
    \vspace{-1em}
\end{table}

\subsection{Additional Study}
\label{sec:perfom_analysis}

\paragraph{Runtime}
We analyze the runtime efficiency of our controllable motion synthesis approach. Each stage of our diffusion model is trained with a diffusion time step $T=50$. \Cref{table:inference_speed} presents the sampling times of each baseline method for comparison.

\begin{figure}[h!]
\centering
\includegraphics[width=0.97\linewidth]{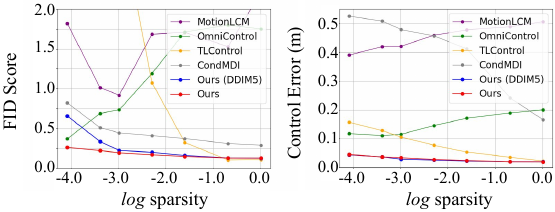}
\vspace{-0.3em}
\caption{We plot the performance of log control sparsity (x-axis) against the \textit{FID} score and \textit{Control error} (y-axis), which assess motion quality and precision, respectively. Our framework maintains consistent performance across varying control input sparsity, outperforming all baselines.}
\label{fig:control_sparsity}
\vspace{-1.2em}
\end{figure}

While MotionLCM~\cite{motionlcm} demonstrates real-time runtime performance, it lacks controllability and produces relatively low-quality motion.
Additionally, we validate that by leveraging the DDIM sampling strategy, our method not only reduces sampling time but also maintains high control precision and motion quality in \cref{tab:experiments_sparse_control} and \cref{fig:control_sparsity}. This allows us to outperform other optimization-based controllable motion synthesis works~\cite{xie2023omnicontrol, wan2023tlcontrol} in terms of runtime efficiency, control accuracy, and overall motion quality.
Note that our decomposed pipeline significantly reduces the runtime compared to other single-stage pipelines (CondMDI, Ours w/o decomp. OmniControl).

\begin{table}[h!]
    \centering
    \resizebox{0.95\columnwidth}{!}{
    \begin{tabular}{ccccc}
        \toprule
        Method & MotionLCM~\cite{motionlcm} & Ours (DDIM 5) & Ours (DDIM 10) & TLControl~\cite{wan2023tlcontrol} 
        \\
        \midrule
        Time (s) & 0.034  & 0.7 & 1.4 & 3.5 \\
        \midrule
        \midrule
        Method & Ours (DDPM) &  CondMDI~\cite{cohan2024flexible} & Ours w/o decomp. & OmniControl~\cite{xie2023omnicontrol} \\
        \midrule
        Time (s) & 7.1 & 39.5 & 41.2 & 152.6 \\
        \bottomrule
    \end{tabular}
    }
    \caption{Time required for motion control.}
    \label{table:inference_speed}
    \vspace{-1.86em}
\end{table}

\paragraph{Robustness to Varying Input Control}

Our controllable motion synthesis framework incorporates a random control mask, $\mathbf{m_C}$, within the key joint trajectory model, ensuring stable performance across diverse selection schemes of control signals.
\Cref{tab:experiments_joint_configuration} presents performance with different sets of control joints.
For the \textit{cross} joint selection scheme, we first sample frames either at uniform intervals of $r$ or with a random probability $p$ then randomly retain 50$\%$ of the keyjoint control signals from the selected frames. This process results in multiple joint signals being selected for each chosen frame.  For the \textit{pelvis} and \textit{right wrist} joint selection schemes, frames are similarly selected at uniform intervals of $r$ or with probability $p$, and only the corresponding joint signals are retained.
The results verify that our method consistently achieves strong performance across varying control signal selection strategies at test time, demonstrating robustness to different frame sampling and joint selection approaches, ranging from single to multi-joint control.

\begin{table}[h!]
    \centering
    \renewcommand{\arraystretch}{1.0}
\resizebox{0.95\linewidth}{!}{
\begin{tabular}{llccccc}
\toprule
\multicolumn{1}{p{1.0cm}}{Joint Select} & \multicolumn{1}{p{1.7cm}}{Frame Select} & \multirow{2}{*}{\centering FID~$\downarrow$} & \multicolumn{1}{p{1.42cm}}{\centering Control Err. ($m$)~$\downarrow$} & \multicolumn{1}{p{1.65cm}}{\centering R-precision (Top-3)~$\uparrow$} & \multirow{2}{*}{\centering Div.~$\rightarrow$} & \multicolumn{1}{p{1.37cm}}{\centering Foot Skating $\downarrow$} \\

\midrule
\rowcolor{gray!15}
- & -      & 0.002 & 0.000 & 0.797 & 9.503 & 0.000 \\

\midrule
\multirow{7}{*}{Cross}
& $r=1$    & 0.127 & 0.019 & 0.681 & 9.518 & 0.071 \\
& $r=2$    & 0.128 & 0.019 & 0.680 & 9.539 & 0.070 \\
& $r=5$   & 0.148 & 0.024 & 0.681 & 9.554 & 0.069 \\
& $r=10$   & 0.171 & 0.027 & 0.678 & 9.402 & 0.074 \\
& $r=20$   & 0.195 & 0.033 & 0.677 & 9.575 & 0.064 \\
& $r=30$   & 0.224 & 0.036 & 0.673 & 9.674 & 0.061 \\
& $r=60$   & 0.263 & 0.044 & 0.659 & 9.627 & 0.062 \\
 \midrule
 \multirow{6}{*}{Cross}
& $p=0.02$  & 0.261 & 0.037 & 0.658 & 9.621 & 0.061 \\
& $p=0.05$   & 0.210 & 0.030 & 0.673 & 9.589 & 0.063 \\
& $p=0.1$   &  0.179 & 0.027 & 0.677 & 9.489 & 0.065 \\
& $p=0.2$   & 0.160 & 0.023 & 0.680 & 9.566 & 0.067 \\
& $p=0.5$   & 0.135 & 0.020 & 0.681 & 9.535 & 0.070 \\
& $p=1.0$   & 0.127 & 0.019 & 0.681 & 9.520 & 0.072 \\
 \midrule
 \multirow{2}{*}{Pelvis}
& $p=0.02$  & 0.245 & 0.039 & 0.656 & 9.681 & 0.059 \\
& $p=0.5$   & 0.243 & 0.023 & 0.675 & 9.583 & 0.073 \\
 \midrule
 \multirow{2}{*}{\shortstack{Right\\Wrist}}
& $p=0.02$  & 0.278 & 0.031 & 0.655 & 9.731 & 0.058 \\
& $p=0.5$   & 0.352 & 0.023 & 0.665 & 9.361 & 0.046 \\
 \bottomrule
\end{tabular}
}
\vspace{-0.6em}
\caption{Quantitative evaluation of diverse control signal selection strategies. Our method demonstrates robust performance, regardless of sparsity, the number or combinations of multiple joints.
}
\label{tab:experiments_joint_configuration}
\vspace{-0.9em}
\end{table}

\Cref{fig:control_sparsity} presents a further analysis  across varying levels of input sparsity.
Our method maintains consistent performance even under highly sparse control signals while synthesizing natural motion with high precision.
Furthermore, even when employing DDIM sampling, our method outperforms all baselines while benefiting from faster sampling speeds compared to our original sampling strategy with full diffusion steps.
In contrast, the performance of baseline methods degrades as control signals become sparser. 
By focusing on low-dimensional keyjoint movements in a separate stage, our model achieves superior controllability and motion quality, regardless of control sparsity.

\paragraph{Experiments on Out-of-Distribution Signal}

We evaluate the robustness of our method to out-of-distribution (OOD) control signals by introducing perturbations into sparse control signals.
Specifically, we select root positions with frame intervals of 20 or 30 as the sparse control, and inject random noise of up to 5 cm to create the OOD samples.
As shown in \Cref{tab:experiments_ood}, our method maintains robust performance and preserves controllability, demonstrating resilience to both in-distribution and OOD data.
Since \textit{FID} and \textit{Diversity} metrics are designed to assess closeness to ground truth data, they are not reported in this setting.

\begin{table}[h!]
    \centering
    \renewcommand{\arraystretch}{1.0}
\resizebox{0.8\linewidth}{!}{
\begin{tabular}{lccc}
\toprule
 \multirow{2}{*}{Control Signal} & \multicolumn{1}{p{1.9cm}}{\centering Control Err. ($m$)~$\downarrow$} & \multicolumn{1}{p{1.65cm}}{\centering R-precision (Top-3)~$\uparrow$}  & \multicolumn{1}{p{1.37cm}}{\centering Foot Skating $\downarrow$} \\

\midrule
$r=20$ (In-data)   &  0.036 & 0.669 &  0.058 \\
$r=20$ (OOD)    &  0.039 & 0.676 &  0.061 \\
$r=30$ (In-data)  & 0.038 & 0.676 & 0.061 \\
$r=30$ (OOD)   &  0.043 & 0.661 &  0.060 \\
 \bottomrule
\end{tabular}
}
\vspace{-0.4em}
\caption{Experiments on Out-of-Distribution control signals. By comparing with in-distribution results (denoted as ``In-data''), our method demonstrates robust performance with out-of-distribution signals.
}
\label{tab:experiments_ood}
\vspace{-1.2em}
\end{table}

\section{Conclusion}
\label{sec:conclusion}

We propose a controllable motion synthesis framework that utilizes sparse and flexible keyjoint inputs. Our decomposed, diffusion-based motion synthesis approach effectively handles highly sparse signals while preserving motion quality. Notably, with only 0.0453$\%$ of the input signals, our model achieves a control error of 0.036$m$ and an \textit{FID} score of 0.224, demonstrating strong adherence to provided constraints while preserving natural movement. We further demonstrate the effectiveness and versatility of our framework in goal-driven motion synthesis across a range of scenarios. Additionally, we formulate keyjoint constraints as various functions—including a time-agnostic control approach that eliminates the need for exact timestamps, allowing for temporally flexible synthesis. We believe that our framework holds significant potential for practical applications in controllable motion synthesis.

Our method is not yet real-time and incorporating the autoregressive sampling while executing motion technique described in \cite{TRUMANS} would require developing a real-time controller. We leave this extension for future work.

\paragraph{Acknowledgements}
This work was supported by Creative-Pioneering Researchers Program through Seoul National University and BK21 FOUR program of the Education and Research Program for Future ICT Pioneers, Seoul National University in 2025. Inwoo Hwang is supported by Hyundai Motor Chung Mong-Koo Foundation.

{
    \small
    \bibliographystyle{ieeenat_fullname}
    \bibliography{main}
}

\clearpage
\clearpage
\maketitlesupplementary

In the supplementary materials, we describe the implementation details (Sec.~\ref{sec:suppl_details}) and present ablation results on different sampling strategy (Sec.~\ref{sec:sampling_ablation}).
Please refer to the supplementary video on our project page for additional qualitative results.

\section{Further Details}
\label{sec:suppl_details}

\subsection{Model Details}

We designed our each model using a diffusion framework based on DDPM~\cite{ho2020denoisingdiffusionprobabilisticmodels}, incorporating the U-Net architecture introduced by \cite{karunratanakul2023gmd}. The training was conducted with the AdamW optimizer~\cite{loshchilov2019decoupledweightdecayregularization}, where we set the learning rate to \(1 \times 10^{-4}\) and applied a weight decay of \(1 \times 10^{-2}\). For inference, we utilized classifier-free guidance with a weight \( \mathbf{w} = 2.5 \). Additional details on the hyperparameters for both the network architecture and diffusion process are summarized in Table~\ref{table:hyperparameters}. Both models were trained using a diffusion process with 50 steps. The key-joint diffusion model is comparatively smaller in size in contrast to the full-body completion diffusion model, which consists of a channel dimension of 128.

\begin{table}[h]
    
    \centering
    \resizebox{0.9\columnwidth}{!}{
    \begin{tabular}{ccc}
        \toprule
        Hyperparameter & Key-Joint Model & Full-Body Model
        \\
        \midrule
        Learning rate & 1e-4  & 1e-4 \\
        Optimizer & Adam W & Adam W\\
        Weight decay & 1e-2 & 1e-2\\
        Batch size & 64  & 64 \\
        Channels dim & 128 & 512 \\
        Channel multipliers & $[2,2,2,2]$ & $[2,2,2,2]$\\
        Variance scheduler & Cosine~\cite{nichol2021improveddenoisingdiffusionprobabilistic} & Cosine~\cite{nichol2021improveddenoisingdiffusionprobabilistic}\\
        Diffusion steps & 50 & 50\\
        Diffusion variance & $\tilde{\beta} = \frac{1-\alpha_{t-1}}{1-\alpha_t}\beta_t$ & $\tilde{\beta} = \frac{1-\alpha_{t-1}}{1-\alpha_t}\beta_t$\\
        EMA weight ($\beta$) & 0.9999 & 0.9999\\
        Guidance weight ($\mathbf{w}$)  & 2.5 & 2.5 \\
        \bottomrule
    \end{tabular}
    }
    \caption{Hyperparameters of each model}
    \label{table:hyperparameters}
\end{table}

\subsection{Baseline Details}

For the baselines OmniControl~\cite{xie2023omnicontrol}, TLControl~\cite{wan2023tlcontrol}, MotionLCM~\cite{motionlcm}, and DNO~\cite{karunratanakul2023dno}, we utilize the officially released checkpoints for evaluation. For CondMDI~\cite{cohan2024flexible}, in order to support arbitrary joint-level control with their strategy, we train the model using global position representations for all joints.

\subsection{Goal-Driven Motion Synthesize Task}

To capture the spatial relationship between body and target locations, we introduce a body shape encoding. This encoding is represented as a continuous shape feature $\mathbf{b}$, which is derived from a set of key measurements. These measurements include joint-to-joint distances obtained from the T-pose, such as:  
$[root, head]$,  
$[left\_shoulder, right\_shoulder]$,  
$[shoulder, wrist]$,  
$[left\_pelvis, right\_pelvis]$,  
$[pelvis, feet]$.  
Additionally, depth measurements for chest and hip thicknesses are computed by evaluating the distances between their front and rear vertices. These measurements collectively form a compact and continuous representation of the body’s proportions.
In order to model various motion dynamics within a unified framework, we assign a unique action label to each task, conditioning the network on these labels.

We evaluate our approach to goal-driven motion synthesis task by training a unified network that integrates multiple scenarios. In each scenario, control signals are provided as an initial pose paired with target control joints at the final frame. Specifically, \textit{reaching target hand positions}~\cite{araujo2023circle} focuses on controlling the right-hand position, \textit{climbing with rock constraints}~\cite{yan2023cimi4d} involves controlling both hands and feet, and \textit{sitting with hand control}~\cite{zhang2022couch} requires controlling both hands at the final frame. Our unified network is trained to address multiple dynamics and control settings simultaneously.

\paragraph{Dataset Description}

We collect a variety of tasks that require control over different target joints at the final frame and involve multiple motion dynamics.
For the scenario of \textit{reaching target hand positions}, we utilize the dataset from \cite{araujo2023circle}. Specifically, we extract sequences of reaching motions and augment them by mirroring the left-hand reaching motions to the right hand, thus generating a total of 3,138 right-hand reaching sequences. To define the goal position, we identify the farthest point reached by the hand from its initial location. In our experimental setup, 2,510 samples are designated for training.
In the case of the \textit{climbing with rock constraints} scenario, we leverage the dataset from \cite{yan2023cimi4d}. We carefully select sequences that depict the subject detaching from one climbing rock and securely reaching for another. This process yields 156 motion samples.
For the \textit{sitting with hand control} scenario, we use the dataset from \cite{zhang2022couch}. We extract 160 sequences that begin with the subject in a stable position and end when they are seated in a chair.
To ensure consistency across all tasks, we standardize the dataset by aligning the subject’s face direction to the $+z$ axis at the initial frame and setting the root position at the origin.

\section{Ablation Study on Sampling Strategy}
\label{sec:sampling_ablation}

We perform a quantitative evaluation of various sampling strategies for diffusion models. Specifically, we train a keyjoint trajectory model and a full-body completion model using 50-step diffusion models and apply both DDPM and DDIM-based diffusion sampling strategies. The results demonstrate that our method maintains consistent performance even with 5-step diffusion sampling and continues to outperform other baselines, achieving high precision and natural motion quality, as supported by Figure 3 of the main paper.

\begin{table}[h!]
    \centering
    \renewcommand{\arraystretch}{1.0}
\resizebox{1.0\linewidth}{!}{
\begin{tabular}{llccccc}
\toprule
\multirow{2}{*}{Sampling} & \multicolumn{1}{p{1.2cm}}{Frame Select} & \multirow{2}{*}{\centering FID~$\downarrow$} & \multicolumn{1}{p{1.42cm}}{\centering Control Err. ($m$)~$\downarrow$} & \multicolumn{1}{p{1.65cm}}{\centering R-precision (Top-3)~$\uparrow$} & \multirow{2}{*}{\centering Div.~$\rightarrow$} & \multicolumn{1}{p{1.37cm}}{\centering Foot Skating $\downarrow$} \\

\midrule
\rowcolor{gray!15}
- & -      & 0.002 & 0.000 & 0.797 & 9.503 & 0.000 \\

\midrule
\multirow{7}{*}{DDPM (50)}
& $r=1$    & 0.127 & 0.019 & 0.681 & 9.518 & 0.071 \\
& $r=2$    & 0.128 & 0.019 & 0.680 & 9.539 & 0.070 \\
& $r=5$   & 0.148 & 0.024 & 0.681 & 9.554 & 0.069 \\
& $r=10$   & 0.171 & 0.027 & 0.678 & 9.402 & 0.074 \\
& $r=20$   & 0.195 & 0.033 & 0.677 & 9.575 & 0.064 \\
& $r=30$   & 0.224 & 0.036 & 0.673 & 9.674 & 0.061 \\
& $r=60$   & 0.263 & 0.044 & 0.659 & 9.627 & 0.062 \\
 \midrule
\multirow{7}{*}{DDIM (10)}
& $r=1$    & 0.136  & 0.019  & 0.678  &9.559  & 0.075 \\
& $r=2$    & 0.141  & 0.020  & 0.681 & 9.574  & 0.074 \\
& $r=5$   &  0.158 & 0.022 & 0.682 & 9.599  & 0.073 \\
& $r=10$   & 0.186  & 0.025  & 0.675 & 9.632 & 0.069 \\
& $r=20$   & 0.222   & 0.030 & 0.681 &  9.644 & 0.067  \\
& $r=30$   & 0.254 & 0.037 & 0.673 & 9.621 & 0.063 \\
& $r=60$   & 0.297  & 0.041  & 0.661 & 9.615  & 0.065 \\
 \midrule
 \multirow{7}{*}{DDIM (5)}
& $r=1$    & 0.127 & 0.019 & 0.678 & 9.528 & 0.072 \\
& $r=2$     &  0.140 & 0.019  & 0.686 & 9.573  & 0.073 \\
& $r=5$    &  0.164  & 0.022  & 0.678  & 9.582   & 0.072 \\
& $r=10$   & 0.202 & 0.025 & 0.674 & 9.543 & 0.065 \\
& $r=20$   & 0.236  & 0.031  & 0.665 & 9.521  & 0.069  \\
& $r=30$   & 0.338 & 0.037 & 0.655 & 9.356 & 0.065 \\
& $r=60$    & 0.658  & 0.045   & 0.622 & 9.147 & 0.073 \\
 \bottomrule
\end{tabular}
}
\caption{Quantitative evaluation of various sampling strategies for diffusion models.
}
\label{tab:sampling_strategy}
\end{table}

\section{Additional Results}
\label{sec:suppl_add_results}

\subsection{Results on Challenging or Unseen Scenarios}

\begin{figure}[h!]
\centering
\includegraphics[width=0.95\linewidth]{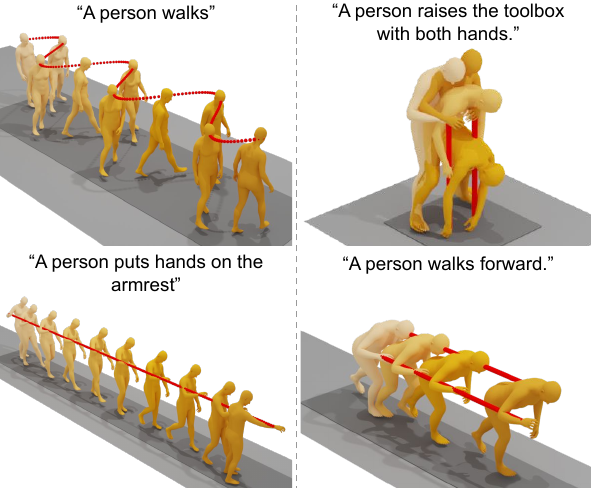}
\caption{Qualitative results on manually specified challenging control signals, such as S-curves and straight paths. Our method successfully follows the intended trajectories, demonstrating strong generalization to unseen and difficult motion constraints.}
\label{fig:unseen}
\end{figure}

We further conduct experiments using manually specified control signals in challenging forms (e.g., S-curves or straight lines), which are unseen and difficult scenarios. 
These results highlight the robustness and generalization capability of our method (Figure~\ref{fig:unseen}).

\subsection{Results on More Expressive Prompts}

\begin{figure}[h!]
\centering
\includegraphics[width=0.95\linewidth]{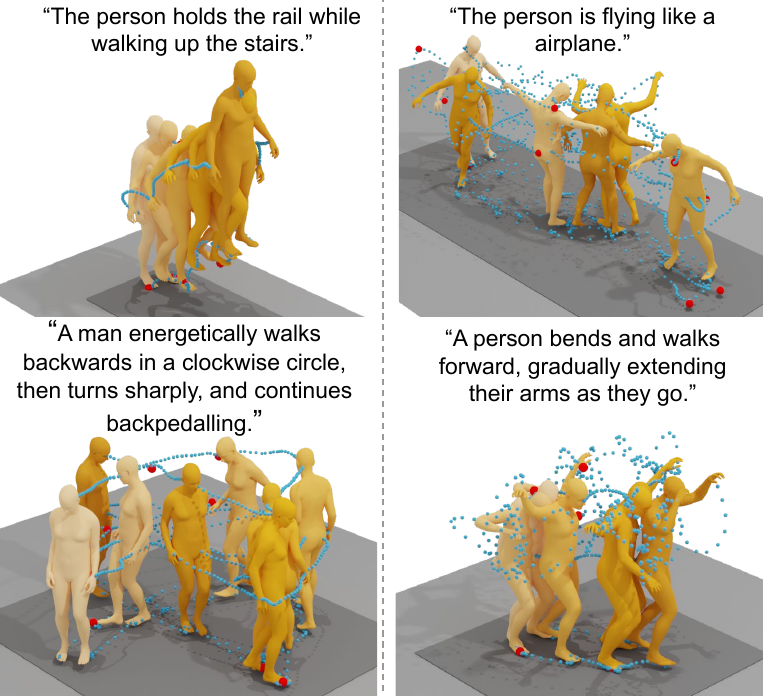}
\caption{Motion generation results for expressive prompts. Our method produces vivid and diverse motions in response to complex textual descriptions while respecting the given sparse control signals.}
\label{fig:expressiven}
\end{figure}

\noindent We provide results for more expressive and complex prompts in Figure~\ref{fig:expressiven}. Our method generates rich, detailed, and vivid motions, demonstrating its ability to capture expressive textual descriptions while faithfully adhering to sparse control signals such as “flying like an airplane,” “extending arms,” “walks backward,” and “walking up the stairs”.

\end{document}